\documentclass{article}

\usepackage{enumitem} 
\usepackage{geometry} 
\usepackage{array} 
\usepackage{listings} 

\usepackage[utf8]{inputenc} 
\usepackage[T1]{fontenc}    
\usepackage{hyperref}       
\usepackage{url}            
\usepackage{booktabs}       
\usepackage{amsfonts}       
\usepackage{nicefrac}       
\usepackage{microtype}      
\usepackage{lipsum}
\usepackage{fancyhdr}       
\usepackage{graphicx}       
\graphicspath{{media/}}     
\usepackage{subfigure}
\title{Casevo: A Cognitive Agents and Social Evolution Simulator}

\author{
  \textbf{Zexun Jiang$^1$ $^2$, Yafang Shi$^1$, Maoxu Li$^1$, Hongjiang Xiao$^2$\thanks{Corresponding author: xiaohj@cuc.edu.cn},} \\
   \textbf{Yunxiao Qin$^2$,Qinglan Wei$^1$, Ye Wang$^2$, Yuan Zhang$^2$} \\
   $^1$School of Data Science and Intelligent Media \\
  Communication University of China \\
  $^2$State Key Laboratory of Media Convergence and Communication\\
  Communication University of China
}

\begin{document}

\maketitle

\begin{abstract}
In this paper, we introduce a multi-agent simulation framework Casevo (Cognitive Agents and Social Evolution Simulator), that integrates large language models (LLMs) to simulate complex social phenomena and decision-making processes. Casevo is designed as a discrete-event simulator driven by agents with features such as Chain of Thoughts (CoT), Retrieval-Augmented Generation (RAG), and Customizable Memory Mechanism. Casevo enables dynamic social modeling, which can support various scenarios such as social network analysis, public opinion dynamics, and behavior prediction in complex social systems. To demonstrate the effectiveness of Casevo, we utilize one of the U.S. 2020 midterm election TV debates as a simulation example. Our results show that Casevo facilitates more realistic and flexible agent interactions, improving the quality of dynamic social phenomena simulation. This work contributes to the field by providing a robust system for studying large-scale, high-fidelity social behaviors with advanced LLM-driven agents, expanding the capabilities of traditional agent-based modeling (ABM). The open-source code repository address of casevo is \url{https://github.com/rgCASS/casevo}.
\end{abstract}


\section{Introduction}

In recent years, with the rapid development of artificial intelligence technology, large language models (LLMs) and agents have shown great potential in various fields. In social science and behavioral research, conducting simulations with agents is one of the commonly used research methods in sociology and communication studies \cite{miller2009complex}. It can help researchers understand complex social phenomena and human behavior \cite{gilbert2005simulation}. Traditional social network simulation systems often rely on agents with pre-defined rules and Mathematical Function Models, making it difficult to capture the dynamics and complexity of human behavior \cite{castelfranchi1998modelling}. In contrast, agents with LLMs are capable of generating highly realistic natural language texts, simulating human decision-making processes and behavioral patterns \cite{bubeck2023sparks}. Therefore, multi-agent simulation can significantly enhance the realism and effectiveness of social simulation. 

In this paper, we introduce Casevo (a Cognitive Agents and Social Evolution Simulator), a multi-agent framework, focusing on the simulation of social interaction and communication based on complex networks. Casevo is built upon the Mesa framework with features for complex and realistic social behaviors and decision-making processes \cite{masad2015mesa}. At the micro level, Casevo supports the construction of highly customized Agents with various features, including Role Insertion, Chain of Thought (CoT), Long-term and Short-term Memory Mechanisms, and so on. These features extend the reasoning capabilities of agents, enabling them to simulate more complex reasoning processes, including event planning and reflection.  At the overall level, Casevo supports customized networks for agent interaction, global events, and also supports the parallel optimization of requests to base LLMs. 

To demonstrate the advantages and potential of Casevo, one of the U.S. 2020 midterm election TV debates is utilized as a simulation example, in which agents represent voters, and reproduce voter behavior through the election process. The simulation demonstrates the effectiveness of Casevo and shows that it may provide a new methodology and tool for future social studies. 

The paper is structured as follows: Section 2 reviews related work.  Section 3 is the overview of Casevo. Section 4 details the module implementation. Section 5 shows the concrete implementation of the election simulation and the analysis of the results. Section 6 discusses the future work. Section 7 summarizes the main contributions of the paper. Through this study, we hope to provide a new perspective for social science and behavioral research, promote the development of multi-agent systems, and explore the application potential of LLMs in a wider range of domains.

\section{Related Work}

This section discusses two key areas most relevant to this research: Multi-Agent Systems and Social Network Analysis. These areas form the core foundation for the development of a social network simulation system.

\subsection{Multi-Agent System (MAS)}
MAS has emerged as a powerful paradigm for simulating and solving complex problems in a variety of domains, including social simulations, robotics, economics, and environmental management. MAS involves multiple autonomous agents, each of which can make decisions and interact with other agents within a shared environment. The interactions between agents are often governed by predefined rules, but the overall system's behavior can emerge as a result of these interactions. MAS has been particularly useful in modeling scenarios where individual agents have diverse goals, strategies, and levels of autonomy \cite{wooldridge1995intelligent}. 

The integration of LLMs into MAS has opened new possibilities for creating more sophisticated agents capable of engaging in complex decision-making, problem-solving, and negotiations. LLMs, with their ability to generate human-like text and handle nuanced contextual information, enable agents to interact in ways that were previously unattainable. For instance, LLM-Debate \cite{du2023improving} introduces the concept of debate, allowing agents to engage with the responses of their peers. When these responses conflict with an agent's judgments, a mental argumentation process unfolds, leading to more refined solutions. GPT-Negotiation \cite{fu2023improving} illustrates how LLMs can autonomously engage in negotiation tasks, demonstrating both strategic reasoning and adaptive learning, a critical advancement for MAS. ChatEval \cite{chan2023chateval} establishes a role-playing multi-agent referee team that assesses the quality of text generated by LLMs through self-initiated debates. This process achieves a level of excellence comparable to that of human reviewers. Corex \cite{sun2023corex} incorporates a variety of collaborative paradigms, including Debate, Review and Retrieve modes, which collectively enhance the factuality, fidelity and reliability of the argumentation process. These paradigms encourage task-agnostic approaches that allow LLMs to think outside the box, effectively counteracting hallucinations and leading to more accurate solutions. 

The application of MAS in social network simulations has proven valuable in capturing the intricacies of human behavior and social dynamics. Recent work by Zhou et al. (2022) \cite{zhou2022game} utilized MAS to simulate multi-agent interactions in a multiplayer game environment, using graph theory and system transformation methods to model the interactions. This study underscored the importance of network structures in MAS and showed how agent behavior can be influenced by network topology, which is central to social network modeling. In parallel, advancements in incorporating emotion and behavior-driven agents, as seen in the humanoid agent models discussed by Wang et al. (2022)[12], provide a more realistic simulation of human-like interactions by embedding psychological elements such as emotions and relational closeness into agent behavior. Systems like WarAgent \cite{hua2023war} demonstrate how MAS can model large-scale international conflicts, simulating complex alliances and responses based on evolving political and social networks. Similarly, S3 \cite{gao2023s} applies MAS to simulate public opinion evolution in social networks, where agents' opinions shift in response to external stimuli and interactions, thus offering insights into collective behavior dynamics. 

Recent developments have introduced new frameworks for combining MAS with LLMs, such as Swarm by OpenAI (2023), a lightweight framework designed to facilitate the integration of multi-agent systems with LLMs. This approach makes it easier to deploy complex multi-agent simulations across a variety of domains, streamlining the orchestration of agents and improving scalability. By leveraging LLMs for agent interaction and decision-making, these systems are capable of simulating more sophisticated social processes, such as negotiation, persuasion, and opinion dynamics.

\subsection{Social network analysis}
Social Network Analysis (SNA) has long been employed as a powerful methodology for studying the structure and dynamics of social networks. SNA traditionally uses graph theory, statistical methods, and machine learning techniques to analyze relationships between nodes and edges within a network, aiming to identify central nodes, community structures, and paths of information flow \cite{wasserman1994social}. For example, Friedkin and Johnsen (1997) \cite{friedkin1997social} \cite{barabasi2013network} explored how social influence processes could be modeled in dynamic networks where the structure of the network itself evolves over time. As Artificial Intelligence (AI) technology continues to develop, social simulation research has also evolved. Vespignani (2012) \cite{vespignani2012modelling} demonstrated how SNA combined with computational models could capture the spread of rumors, diseases, or even misinformation in a more realistic and actionable way. However, these traditional approaches often rely on static data, which limits their ability to adapt to the dynamic, evolving nature of social networks. In the context of dynamic social behaviors—such as opinion shifts, rumor propagation, or public sentiment—traditional SNA is often insufficient. 

Recent advancements have sought to address this limitation by integrating SNA with agent-based modeling (ABM) and multi-agent systems (MAS), where individual agents exhibit dynamic behaviors that evolve over time. This approach allows for the representation of not just the static structure of networks but also the dynamic, often unpredictable interactions among agents. It enables a more nuanced understanding of how networks function and evolve.  For example, in social network simulations, agents' behaviors, such as communication, collaboration, and conflict resolution, can now be modeled dynamically, taking into account both network structure and individual agent behavior \cite{gao2016universal} \cite{newman2018networks}. Alvim M.S. and Amorim B. (2023) \cite{alvim2023formal} proposed a polarization model based on confirmation bias, analyzing how users gradually form extreme opinions within social networks. Their study demonstrated that the degree of polarization is influenced by both the network topology and the initial distribution of users' opinions. Similarly, Vilone D. and Polizzi E. (2024) \cite{vilone2024modeling} using a multi-agent simulation model, explored opinion misunderstandings and the spiral of silence within various network topologies. They highlighted that highly connected nodes in the network, such as opinion leaders, may play a critical role in shaping information dissemination. Systems like Generative agents \cite{park2023generative} leverage agent-based simulations to emulate human behavior in social settings, allowing for the generation of realistic, dynamic interactions that traditional SNA methods could not address. Furthermore, applications such as WarAgent and S3 \cite{hua2023war} \cite{gao2023s} have shown the potential of these hybrid models in simulating large-scale social phenomena like international conflicts and public opinion evolution, further proving the robustness of combining SNA with agent-based modeling.

\section{Overview}

\begin{figure}
  \centering
  \includegraphics[width=12cm]{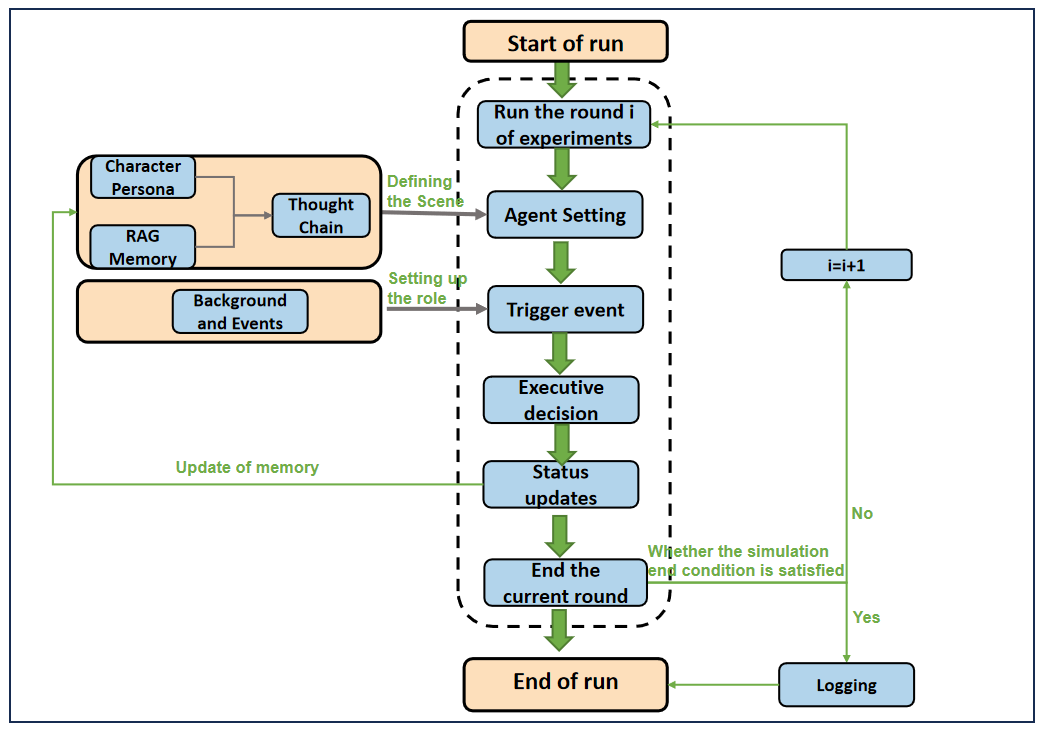}
  \caption{Schematic implementation of Casevo rounds.}
  \label{fig:fig1}
\end{figure}

Compared to existing open-source MASs, Casevo is specifically designed and built for simulating social interactions on complex networks. Casevo is built upon the Mesa framework \cite{masad2015mesa}, designed for discrete-event simulation, using a round-robin updating mechanism to manage the behavior of agents and event scheduling. While the Mesa framework has been widely used in the field of ABM, Casevo extends features as complex semantic reasoning and social network interactions for modeling of complex social phenomena such as election simulation and behavior prediction.

Figure \ref{fig:fig1} shows a typical example of Casevo’s round-based execution. In each round, all agents update their decisions and perform the corresponding behaviors based on the current state, environmental inputs, and interactions with other agents. The turn-by-turn updating ensures the synchronization of system behaviors and the orderly scheduling of events, which is particularly suitable for social dynamic simulation scenarios that need to progress step-by-step.

\begin{figure}
  \centering
  \includegraphics[width=8cm]{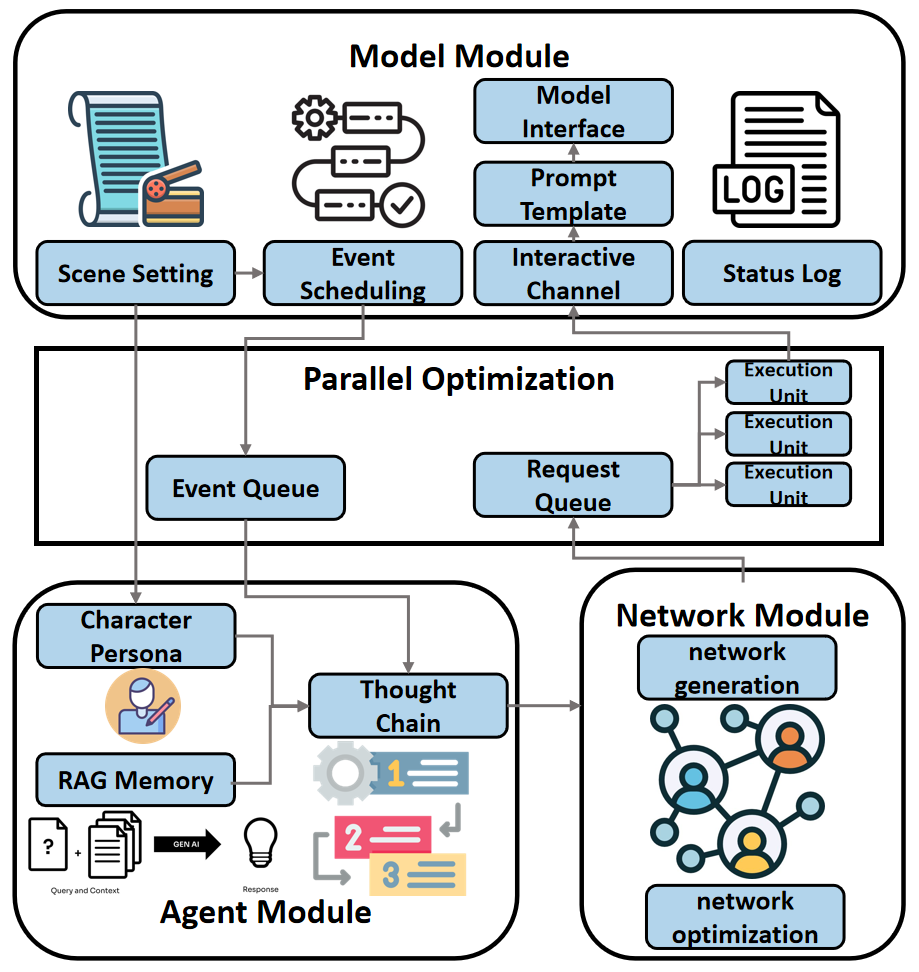}
  \caption{The architecture of the system.}
  \label{fig:fig2}
\end{figure}

As shown in Figure \ref{fig:fig2}, Casevo utilizes a modular design manner and integrates the functions of agent behavior generation, memory management, network interaction, and parallel processing to form a highly flexible and scalable system. The system is capable of customizing the simulation scenarios according to the user's needs, including the personality settings of the agent, the network topology, and the intervention of external events. These customization features make Casevo applicable to a wide range of research areas, including social network analysis, the study of public opinion dynamics, the prediction of behavior in complex systems, and the simulation of decision-making processes in collective environments.

The architecture of Casevo is composed of 4 key modules: the Model Module, the Agent Module, the Parallel Optimization Module, and the Network Module. These components work together to support the simulation of complex systems, enabling high-level interactions between agents while efficiently managing large-scale computations. Each module plays a distinct role in managing agent behaviors, decision-making processes, memory recall, and network interactions, ensuring a seamless and efficient simulation flow.

\section{Module Implementation}

Through the design and integration of four modules, Casevo achieves comprehensive simulation from autonomous decision-making of agents to complex social networks. It provides an efficient and flexible platform for the research and exploration of a wide range of complex social phenomena.

\subsection{The Model Module}
The Model Module serves to set up the environment where the simulation takes place. It is responsible for scenario settings, event scheduling, interaction channels, prompt templates, and model interfaces. In addition, it handles state logging, ensuring that all changes and actions are recorded throughout the simulation.

\begin{itemize}
\item Scenario Settings: Casevo reads the configuration parameters provided by the users, defining the initial environment of the simulation. This includes determining the number of agents, their roles, and other key parameters necessary for the simulation.
\item Event Scheduling: The Model Module controls the scheduling of events within the simulation. It ensures that interactions between agents occur on time, adhering to the logical flow defined by the round-based update mechanism.
\item Interaction Channels: Agents communicate through well-defined interaction channels, allowing them to share information, influence each other, and adapt their behavior accordingly. These channels replicate real-world communication pathways, such as social media, public debates, or direct conversations.
\item LLM Interfaces: The integration of LLMs allows agents to utilize multiple models and prompt templates for generating context-aware responses. These templates ensure consistency in agent communication, while the model interfaces support the interaction between the simulation environment and the LLMs, facilitating intelligent decision-making.
\item State Logging: Throughout the simulation, the Model Module logs all changes in agent states, including decisions, interactions, and environmental changes. This data is crucial for post-simulation analysis, providing insights into how agents evolve and interact over time.
\end{itemize}

\subsection{Agent Module}
The Agent Module is responsible for defining the behavior of each agent, managing their memory, and enabling multi-step reasoning through the CoT mechanism. The agents in the simulation are highly customizable, with individual attributes, beliefs, and goals.

\begin{itemize}
\item Role Definition: Each agent is assigned a role, which governs its decision-making process and behavior patterns. These roles are often based on real-world entities, such as voters in an election or users in a social network, ensuring that agents exhibit realistic behaviors aligned with their defined attributes.
\item RAG Memory: The Retrieval-Augmented Generation (RAG) memory system allows agents to recall past interactions and decisions, enabling more nuanced and context-aware behavior generation. This memory mechanism makes it possible for agents to retain information across multiple rounds of interaction, simulating human-like memory and long-term decision-making.
\item CoT: The Mechanism provides agents with the ability to engage in multi-step reasoning. This mechanism allows agents to think through complex decisions, considering multiple factors before concluding. It is crucial for simulating strategic behaviors, such as planning, negotiation, and coalition-building.
\end{itemize}

\subsection{Parallel Optimization Module}
As the complexity of the simulation scale increases, the efficient computation becomes more and more urgent. The Parallel Optimization Module addresses this challenge by managing event and request queues, ensuring that agent actions are processed efficiently.

\begin{itemize}
\item Event Queue: The event queue manages all scheduled events within the simulation. Events can be triggered externally, such as through media broadcasts or social events, or internally, through agent interactions. The system ensures that events are processed in an organized manner, maintaining the flow of the simulation.
\item Request Queue and Execution Units: When agents submit behavior requests, these are placed in the request queue, which is then processed by the system's execution units. By distributing the workload across multiple execution units, the system can handle parallel agent behaviors, ensuring that even large-scale simulations are processed on time.
\end{itemize}

\subsection{Network Module}
The Network Module is responsible for creating and managing the network structure within which agents operate. This module handles both the generation of the network at the start of the simulation and the dynamic adjustments that occur based on agent interactions.

\begin{itemize}
\item Network Generation: Casevo can generate social network structures based on the configuration parameters provided by users. These networks can take on different forms, such as small-world networks or random networks, depending on the type of social system being modeled.
\item Network Optimization: Throughout the simulation, the network structure is dynamically adjusted based on agent interactions. For example, if two agents engage in frequent communication, the strength of their connection in the network may increase, reflecting their growing influence on each other. This dynamic adjustment ensures that the network structure evolves in a manner that mirrors real-world social networks.
\end{itemize}

\subsection{Key Features}
The system architecture of Casevo provides a robust foundation for multi-agent simulation, enabling the flexibility to meet diverse scenario requirements. Building on this architecture, a range of features further enhances its applicability and efficiency in simulating complex social behaviors. 

\begin{enumerate}[label=\alph*.] 
\item Complex Network Support: Casevo provides powerful network generation and optimization capabilities, establishing flexible network structures for agent interactions. This structure not only simulates individual interactions within social networks but also dynamically adjusts agent relationships, reflecting the evolving characteristics of real-world social networks. This allows users to simulate diverse social systems, such as information dissemination in social media or social influence models.
\item RAG Memory Mechanism: The system introduces a Retrieval-Augmented Generation (RAG) memory module, enabling agents to store and recall past events, decisions, and interaction histories. This memory mechanism allows agents to not only make decisions based on the current environment in each round but also adjust their actions based on historical data. This feature enables agents to simulate human-like long-term strategic thinking and memory-dependent decision-making.
\item Round-Based Updates: The system follows Mesa's round-based update model. In each round, agents simultaneously update their states and execute actions. This ordered round progression ensures the efficient handling of complex event scheduling and behavior synchronization across multiple agents.
\item Parallel Optimization: To improve simulation efficiency, Casevo incorporates parallel processing techniques into its execution engine. Through parallel optimization, the system can handle multiple agents' behaviors and decisions simultaneously, improving the overall simulation’s performance. This feature makes Casevo particularly suitable for large-scale social system simulations, such as modeling voter behavior in elections and public opinion dissemination.
\end{enumerate}

Through these enhancements, Casevo maintains the flexibility of the Mesa framework while enhancing its ability to handle complex semantic scenarios and large-scale network interactions. It stands as a powerful and versatile tool for multi-agent simulations. Later we will describe in more detail how Casevo achieves these features through a modular design and the specific functions and interactions of each module.

\section{Scenario Example}

One of the U.S. 2020 midterm election TV debates is utilized as an example to demonstrate the Casevo framework. 

\subsection{Scenario Settings}
This election simulation is based on the 2020 US presidential election and simulates one of the TV public debates between candidates Trump and Biden. Agents play the role of voters to watch the debate, generate ideas or viewpoints based on their personas, and then conduct simulated voting. In addition, agents can interact and share their thoughts with each other, simulating the exchange of opinions and interactions between voters in reality. 

\begin{figure}
  \centering
  \includegraphics[width=8cm]{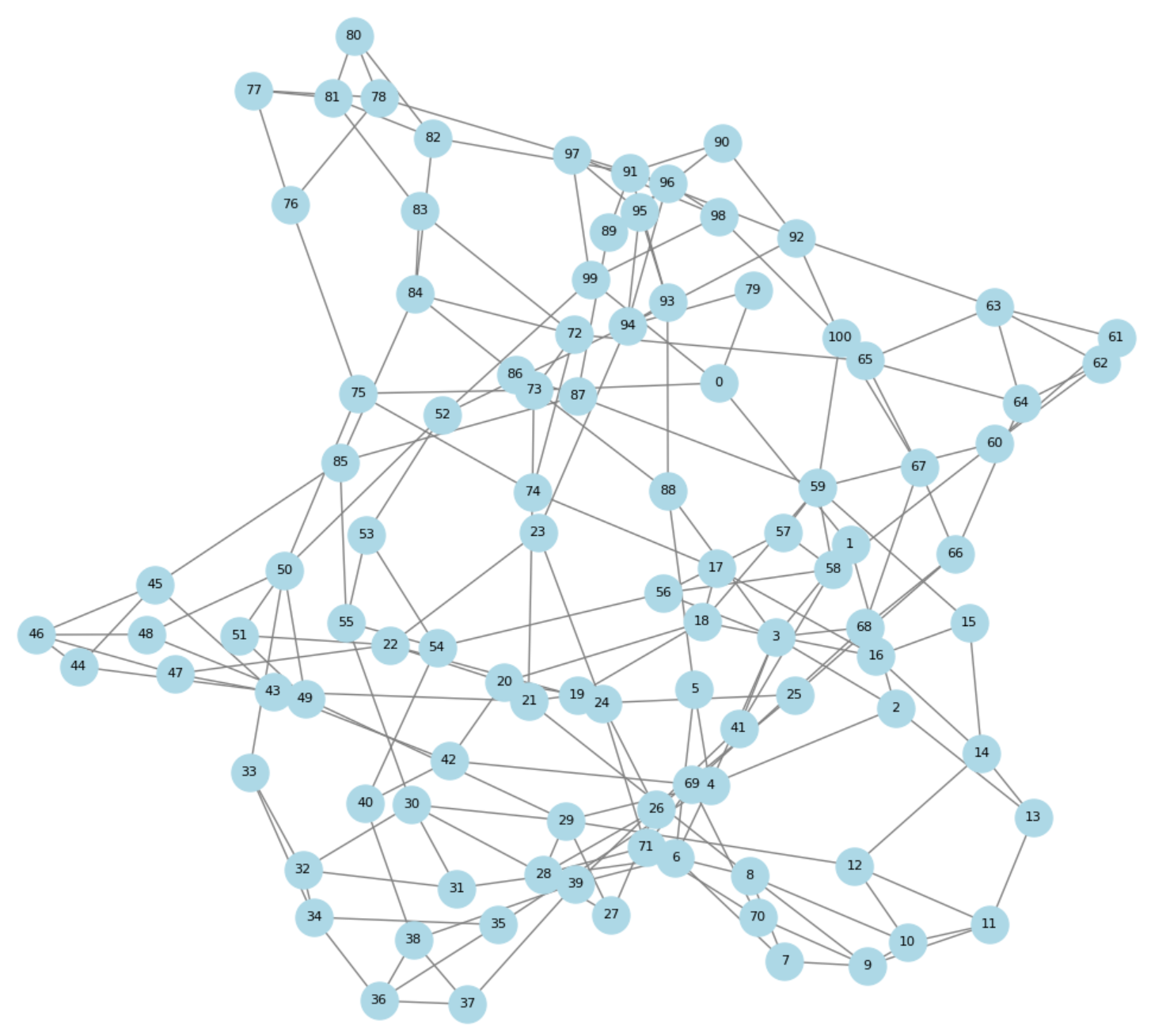}
  \caption{The structure of the network between voters.}
  \label{fig:fig3}
\end{figure}

To simulate the social interaction between voters, the social network structure is a small-world network with 101 nodes, as shown in Figure \ref{fig:fig3}. Each node represents an agent or a voter. The edge between two nodes represents the friendship between the corresponding agents.  The network allows agents to interact with each other, share their opinions, and adjust their voting preferences based on the content of the exchanges.

\subsection{Agent Configuration}
In the election simulation, the configuration of agents is crucial to ensuring that each simulated voter accurately reflects the complex psychology and behavior patterns of real voters in both behavior and decision-making. Each agent represents an individual who evaluates the candidates in the debate based on their characteristics and areas of concern, making voting decisions based on the evaluation results. To describe an individual voter, the main parameters are as follows:

\begin{enumerate}[label=\alph*.] 
\item General description: The basic information of voters, including age, ethnicity, place of residence, etc.
\item Voter Characteristics: Includes background and topics of interest. Each voter agent has different areas of concern that shape their views on various national issues. For instance, some voters may advocate reducing government intervention in society, while others may insist on increasing it.
\end{enumerate}

In order to simulate the real voter situation of the US presidential election, agents are configured with a study on the political typology of American voters \cite{pew_research}. 
It offers a nuanced classification beyond the simple Democratic-Republican division, identifying distinct groups such as steadfast conservatives, stable liberals, and young outsiders, each with unique political behaviors and attitudes. The agents in the experiment are categorized into multiple types based on the main characteristics of party voters in the political typology, with each type representing the typical characteristics of certain voter groups in reality, as shown in Table \ref{tab:table1}. This configuration allows the simulation to encompass voters from different age groups, social backgrounds, and political standings, and to model dynamic changes during the election process through these diverse settings.

\begin{table}
 \caption{Voter Settings}
  \centering
  \begin{tabular}{cc}
    \toprule
    Category & General Characteristics \\
    \midrule
    Conservative White Male & 55 years old, rural community, conservative \\
    Financially Stable Educated White Male & Over 50, financially stable, educated white male \\
    Low-Income, Low-Education White Female & Rural, low-income, low-education white female \\
    Young, Diverse Group & Under 50, young, diverse \\
    Low-Income White Female & Low-income, low education level white female \\
    Financially Struggling Young Liberal Voter & Financially struggling, young, liberal \\
    Elderly Religious Female & Elderly, deeply religious, low education level \\
    High-Income, Highly Educated White & High-income family, highly educated white \\
    Highly Educated Liberal Young White & Highly educated, group-oriented, young liberal white \\
    \bottomrule
  \end{tabular}
  \label{tab:table1}
\end{table}

Following the pre-defined agent configuration, the system generates voter agents that meet the role characteristics and quantity requirements through a rigorous initialization process. Table \ref{tab:table2} is an example of agent profiles.

\begin{table}[ht]
  \centering
  \caption{Profile Example}
  \begin{tabular}{|p{0.9\textwidth}|} 
    \hline
    \textbf{Name:} Alex Thompson \\ 
    \textbf{Age:} 55 years \\ 
    \textbf{Background Description:} A 55-year-old conservative white male living in a rural community. \\
    \textbf{Topics of Interest:} As a devout Christian, he believes religion is essential, supports a greater role for religion in public life, and feels government policies should uphold religious values. He actively participates in government and public affairs, advocating for reduced government intervention in society. On issues such as abortion and same-sex marriage, he holds conservative views, believing that the legalization of same-sex marriage harms the nation’s interests. He firmly believes that being white does not confer any advantage based on race and that significant discrimination exists against white people in contemporary society. On international affairs, he believes that a strong national military is essential and supports military expansion, viewing military strength rather than diplomacy as the best means to ensure peace. \\
    \hline
  \end{tabular}
  \label{tab:table2}
\end{table}

\subsection{Simulation Steps}
Since the presidential debate covers 6 main issues, the simulation is divided into 6 rounds. The main steps for agents in each round include: Watching and Evaluating, Discussing and Interacting, Reflecting and Voting.

\subsubsection{Watching and Evaluating}
At the beginning of each round, candidates broadcast their policy ideas and campaign concepts to voters through the TV debate. At this stage, all agents receive the corresponding transcripts of the debate simultaneously. Each voter processes this information according to his or her inclinations and personality traits to form an initial impression of the candidate as follows.

\begin{enumerate}[label=\alph*.] 
\item Parse the candidates' debate speech text and extract the information so that agents can analyze and assess the candidates' viewpoints.
\item Based on the agents’ interests and the content of the candidates’ speeches, the agents evaluate the candidates.
\item Based on the results of the evaluation, the agents determine their level of support for each candidate.
\end{enumerate}

At this point, agents have not yet made a final voting decision. They are only forming a tentative opinion about candidates. To demonstrate the information processing and decision-making process of the voter agent in the election simulation, Table \ref{tab:table3} shows an example of the detailed content output generated by an agent. 

\begin{table}[ht]
  \centering
  \caption{Example of the Watch and Evaluate Stage}
  \begin{tabular}{|p{\textwidth}|}
    \hline
        "ts": 0, \\
        "owner": "agent\_60", \\
        "type": "listen", \\
        "item": \{ \\
        \quad "source": "public", \\
        \quad "content": "2020 U.S. Presidential Election ......", \\
        \quad "opinion": " \\
        1. Fighting the new crown epidemic: supports Trump's anti-epidemic measures, including closing the border, promoting the use of masks, and accelerating vaccine research and development, but agrees with Biden on the need for increased testing and national standards for epidemic prevention. \par
        2. Economy and jobs: endorses Trump's economic policies, especially tax cuts and measures to promote energy independence, while also supporting Biden's advocacy for more economic assistance for those affected by the epidemic. \par
        3. Healthcare: Supports Trump's push for healthcare reform, including lowering drug prices and promoting Medicaid cooperation across state lines, but also endorses Biden's proposal to expand healthcare coverage. \par
        4. Race relations: supports Trump's efforts on criminal justice reform and police department reform, while agreeing with Biden's response to systemic racism that police reform needs to go further. \par
        5. Environment and energy: supports Trump's energy independence policies and environmental achievements, such as withdrawing from the Paris climate agreement, but agrees with Biden on the need to develop clean energy and reduce greenhouse gas emissions. \par
        6. Foreign policy: recognizing Trump's achievements in diplomacy, such as diplomatic relations with North Korea and the Middle East peace process, but at the same time endorsing Biden's advocacy for greater global cooperation to address international challenges." \\
        \} \\
    \hline
  \end{tabular}
  \label{tab:table3}
\end{table}

\subsubsection{Discussing and Interacting}

Agents interact and exchange opinions with their neighbors (other agents in their social circle) through social networks. Each agent will not only share their opinions about the debate content, but will also receive ideas and feedback from others. 

\begin{table}[ht]
  \centering
  \caption{Examples of Opinion Expressions in Voter Interaction}
  \begin{tabular}{|p{0.9\textwidth}|} 
    \hline
    \textbf{Epidemic response:} supports Trump's anti-epidemic measures (e.g., closing the border, vaccine development) but agrees with Biden's point about increased testing and national standards for prevention. \\
    \textbf{ECONOMY \& JOBS:} Endorses Trump's tax cuts and promotion of energy independence, but agrees with Biden's assertion of the need for financial assistance for those affected by the outbreak. \\
    \textbf{Healthcare:} supports Trump's health care reform, but believes it needs to be available to more people, endorses some of Biden's proposals. \\ 
    \textbf{Race relations:} supports Trump's efforts on justice reform, while agreeing with Biden's stance on tackling systemic racism. \\ 
    \textbf{ENVIRONMENT \& ENERGY:} Supports Trump's energy independence policy while endorsing Biden's proposal to develop clean energy. \\ 
    \textbf{Foreign policy:} supports Trump's foreign policy accomplishments, but sees the need for greater global cooperation to address international challenges, partially agreeing with Biden. \\ 
    \hline
  \end{tabular}
  \label{tab:table4}
\end{table}

As shown in Table \ref{tab:table4}, in the first round Agent No.60 and Agent No.62 generate an exchange in which Agent No.60 expresses his or her views on the six main topics of the debate to Agent No.62, and the content of the exchange is recorded in the voter's memory, and then that voter revisits the two candidates in light of the content of the discussion, generating a new viewpoint, as detailed in Table \ref{tab:table5}.

\begin{table}[ht]
  \centering
  \caption{Example of post-interaction reflection}
  \begin{tabular}{|p{\textwidth}|}
    \hline
        "agree": \{ \\
        \quad "Trump": \{ \\
        \quad\quad "Identity": 0.6, \\
        \quad\quad "Overall view": "Generally supportive, especially on tax cuts, deregulation, and energy independence, but has concerns about its health care reform and neglect of climate change." \\
        \quad \}, \\
        \quad "Biden": \{ \\
        \quad\quad "Identity": 0.4, \\
        \quad\quad "Overall View": "There is consensus on issues such as economic aid, police reform, and clean energy, but disagreement on immigration policy, which is partially in favor of its policies." \\
        \quad \} \\
        \} \\
    \hline
  \end{tabular}
  \label{tab:table5}
\end{table}

\subsubsection{Reflecting and Voting}
At the end of each round, agents entered a reflective phase, as shown in Table \ref{tab:table6}. At this point, each agent retrieves and reflects on its short-term memory based on the previous information and the discussion content, and further adjusts its views by combining them with past historical experiences (e.g., previous candidate performance or past policies). Through this built-in reflective mechanism, agents not only rely on the current discussion, but also synthesize their experiences and memories over time. This stage helps agents make more solid and rational decisions, especially when they are confronted with complex social issues or policy controversies.

\begin{table}[ht]
  \centering
  \caption{Example of a reflection phase}
  \begin{tabular}{|p{\textwidth}|}
    \hline
        "item": \{ \\
        \quad "ori\_opinion": null, \\
        \quad "new\_opinion": "1. TRUMP: I support President Trump's positions on many key issues, including religious liberty, reducing the size of government, and the importance of national defense. At the same time, I agree with the steps he has taken on economic policy, health care, race relations, environment and energy, and foreign policy. Despite my reservations about some of his statements and actions, I still believe he is the candidate who better represents my views. \\
        Biden: While I have reservations about Vice President Biden's views on some issues, such as his attitude toward illegal immigration and his views on public assistance, I also believe that he has put forward sound programs in other areas such as health care, race relations and environmental policy. However, I am concerned about his tendency to support same-sex marriage and expand government intervention. Overall, I think Vice President Biden differs significantly from my views in some areas, but in others there is some agreement." \\
        \} \\
    \hline
  \end{tabular}
  \label{tab:table6}
\end{table}

After reflection, agents make their final voting choices based on the results of the reflection. Each agent's voting decision is directly influenced by the previous phases, including the effectiveness of the candidate's speech, the flow of opinions in the social network, and their own reflective process. The system records each voter's voting choice, providing data for status updates in subsequent rounds and for the final results tally.

\begin{table}[ht]
  \centering
  \caption{Example of a Voting Stage}
  \begin{tabular}{|p{\textwidth}|}
    \hline
        "ts": 0, \\
        "owner": "agent\_60", \\
        "type": "vote", \\
        "item": \{ \\
        \quad "Trump": 0.8, \\
        \quad "Biden": -0.5 \\
        \}, \\
        "owner\_type": 0 \\
    \hline
  \end{tabular}
  \label{tab:table7}
\end{table}

The election example demonstrates how Casevo's modular design can be useful in the simulation of complex social phenomena. By coordinating agent behavior, dynamically adapting social networks, and employing advanced memory and reasoning mechanisms, the system provides a powerful tool for exploring the election process in depth, demonstrating how individual decisions can affect collective outcomes.

\subsection{Evaluation}

\begin{figure}
    \centering
    \subfigure[Voting Result]{
        \includegraphics[width=9cm]{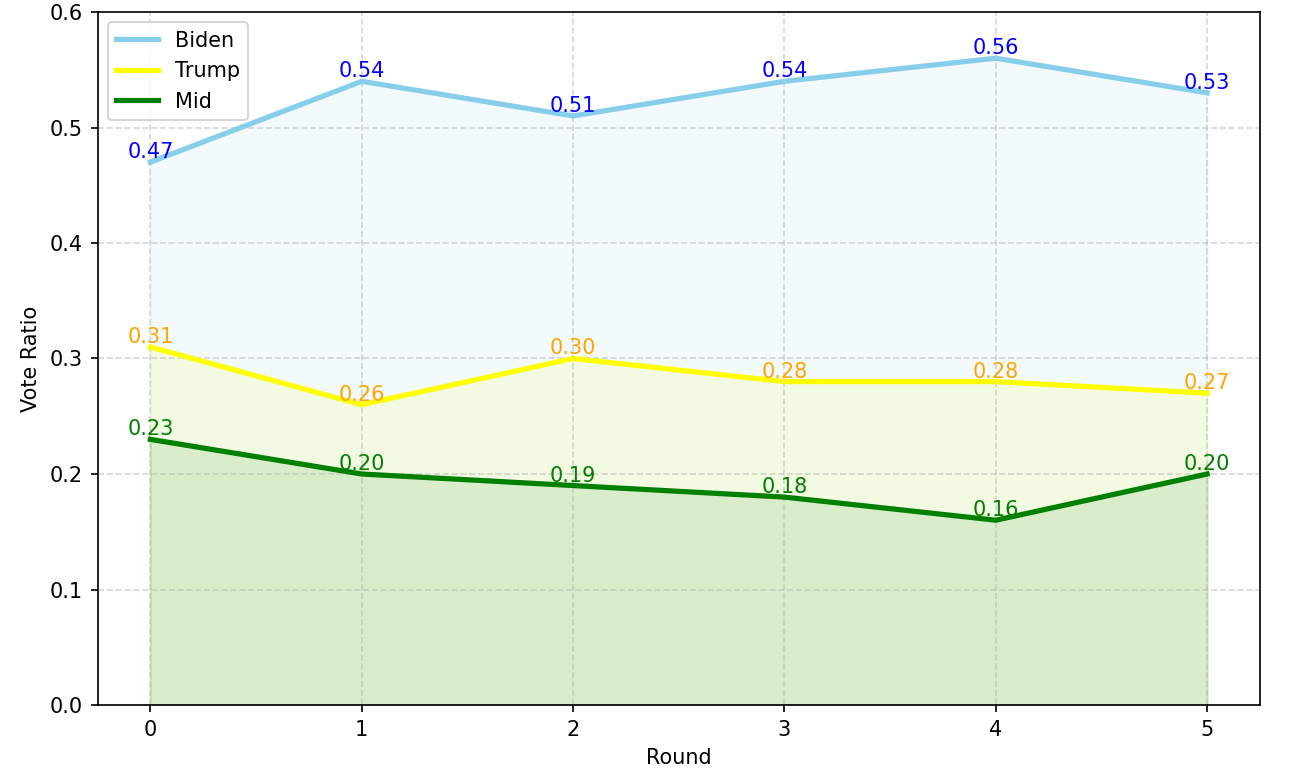}
        \label{fig:fig4}
    }
    \subfigure[Word cloud of voter opinions]{
        \includegraphics[width=5cm]{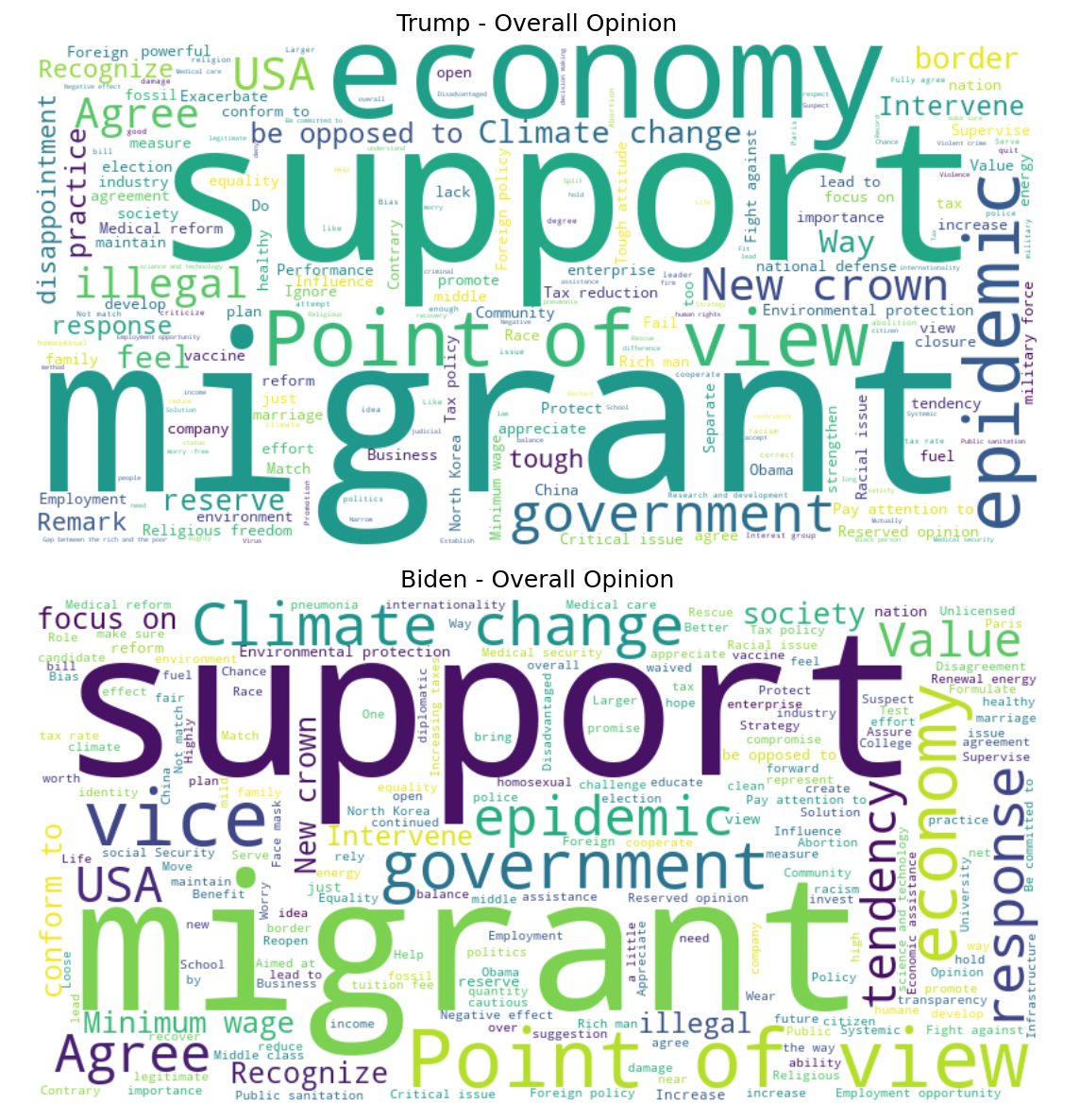}
        \label{fig:fig5}
    }
    \caption{Simulation Result}
    \label{fig:subfigure_example1}
\end{figure}

Figure \ref{fig:fig4} shows the changes in the agents' attitudes toward the candidates and their voting behaviors in each round. It can help researchers understand the dynamic evolution process of voters' opinions.

As shown in Figure \ref{fig:fig4}, Biden's approval rating is relatively high in the initial stages and steadily increases as the election progresses, eventually stabilizing at a high level. Trump's support, on the other hand, shows a downward trend at the beginning of the election, and despite a rebound in some rounds, the overall support fails to pick up significantly. The proportion of neutral voters decreases, suggesting that more and more neutral voters begin to favor either Biden or Trump as the election progressed, but more neutral voters turns to Biden. 

This result suggests that Biden's policies and speeches won over more voters during the simulation, especially in the mid-to-late-debate period, when Biden tilts neutral voters towards his camp through his sustained policy advocacy and communication. Trump, on the other hand, fails to effectively expand his base of supporters, and his support fluctuates and eventually declines slightly.

As shown in Figure \ref{fig:fig5}, opinions about Biden are dominated by positive words such as "support" and "approve", reflecting Biden's strong support among voters. Voters highly approve of Biden's stance on several policy issues, especially on issues such as the response to the new crown epidemic, environmental and energy policies, and race relations, where Biden's policy ideas are consistent with voters' expectations and have gained widespread support. 

By contrast, Opinions about Trump are almost evenly split between positive and negative terms. This suggests that voters are sharply divided on Trump's positions, particularly on key issues such as the response to the epidemic, race relations, and foreign policy, where voters are more antagonistic in their views. A portion of voters hold a favorable view of his policies, while another portion is strongly opposed to them. Overall, Trump's support in the simulation is relatively weak, and his governance strategy fails to win the approval of a wide range of voters. 

The experimental results clearly show the distribution of voters' support for the two candidates and their dynamics through the graphs of changes in support and word cloud maps of voters' perceptions. Biden shows a stronger support tendency in the simulation, while Trump's support is more fluctuating, and these results provide an important data basis for the simulation study of election dynamics.

\section{Discussion and Future Work}
In the previous sections, we presented in detail the system architecture, module design, and specific applications of Casevo in election simulation. Casevo enables a high-quality social simulation and demonstrates its power in complex scenarios. Firstly, in terms of ease of use, the system allows users to easily configure agents and simulation scenarios, avoiding the tedious task of writing complex code. Users can generate complex simulation scenarios by simply entering the appropriate parameters according to their needs, greatly reducing the learning threshold and operational difficulty. Secondly, in terms of efficiency, Casevo is optimized for the parallel behavior of large-scale agents, especially in complex interaction and decision processing. By using parallelization techniques, the system can process a large number of tasks in a short time, greatly improving simulation efficiency. In addition, Casevo has excellent real-time performance, providing instant feedback on the dynamic behavior and decision results of the agent during the simulation process, allowing users to monitor and understand the progress of the simulation in real-time. Together, these features provide a powerful and efficient simulation tool, and the highly modular design of the Casevo architecture allows the system to be easily adapted to different application scenarios.

In future work, we consider introducing more transparent behavioral models or incorporating more logical reasoning mechanisms to enhance the interpretability of agent behavior. At the same time, we explore more effective data integration methods to ensure that the system can better reflect the dynamic processes in the real world; and try to use larger scale and more realistic social network data, as well as more detailed individual behavioral data, to further validate the effectiveness and universality. In addition, besides election simulation, the Casevo architecture can be applied to other simulation scenarios of social dynamics. For example, news dissemination, opinion generation, and market behavior. We plan to integrate more social science theories and behavioral models into Casevo through modular design, so as to enhance its performance capability in different application scenarios.

\section{Conclusion}
In this paper, we propose the Casevo framework, which can help implement a multi-agent system capable of modeling complex social dynamics by integrating LLMs as a decision center. We demonstrate the application of the Casevo in election simulation by creating a large number of voter agents to reproduce the information dissemination, interaction and decision-making behavior during the election process. The results show that the Casevo system is effective in capturing the complexity of voter behaviors and provides new perspectives and tools for election research. Casevo demonstrates strong flexibility and scalability in simulation applications.  Future research directions include optimizing the performance of LLMs, improving the quality of generated content, and exploring more effective methods for real-world data integration. In addition, we will extend the application scenarios of Casevo to further validate its broad applicability in social science research.

\bibliographystyle{unsrt}  
\bibliography{references}

\end{document}